\documentclass[journal]{IEEEtran}

\ifCLASSINFOpdf
\else
   \usepackage[dvips]{graphicx}
\fi
\usepackage{url}
\usepackage{bbding}
\usepackage{graphicx}
\usepackage{float}
\usepackage{amsfonts}
\usepackage{subfigure}
\usepackage{multirow}
\usepackage{amsmath}
\begin{document}

\title{Multi-layer Feature Fusion Convolutional Network for Audio-visual Speech Enhancement}

\author{Xinmeng Xu, Jianjun Hao
\thanks{(\textit{Corresponding author: Jianjun Hao}) }
\thanks{Xinmeng Xu is with the Electronic $\&$ Elect. Engineering, School of Engineering, Trinity College Dublin, Dublin, Ireland (e-mail: xux3@tcd.ie).}
\thanks{Jianjun Hao is with School of Foreign languages, HBUCM, Wuhan, China (e-mail: loyalcolin@163.com).}}

\maketitle

\begin{abstract}
Speech enhancement can potentially benefit from the visual information from the target speaker, such as lip movement and facial expressions, because the visual aspect of speech is essentially unaffected by acoustic environment. In this paper, we address the problem of enhancing corrupted speech signal from videos by using audio-visual (AV) neural processing. Most of recent AV speech enhancement approaches separately process the acoustic and visual features and fuse them via a simple concatenation operation. Although this strategy is convenient and easy to implement, it comes with two major drawbacks: 1) evidence in speech perception suggests that in humans the AV integration occurs at a very early stage, in contrast to previous models that process the two modalities separately at early stage and combine them only at a later stage, thus making the system less robust, and 2) a simple concatenation does not allow to control how the information from the acoustic and the visual modalities is treated. To overcome these drawbacks, we propose a multi-layer feature fusion convolution network (MFFCN), which separately process acoustic and visual modalities for preserving each modality features while fusing both modalities features layer by layer in encoding phase for enjoying the human AV speech perception. In addition, considering the balance between the two modalities, we design channel and spectral attention mechanisms to provide additional flexibility in dealing with different types of information expanding the representational ability of the convolution neural network. Experimental results show that the proposed MFFCN demonstrates the performance of the network superior to the state-of-the-art models.
\end{abstract}


\IEEEpeerreviewmaketitle

\section{Introduction}\label{sec1}
\IEEEPARstart{B}{ackground} noises greatly reduce the quality and intelligibility of the speech signal, limiting the performance of speech-related applications, such as speech recognition \cite{ref3, ref4}, speaker verification \cite{ref5, ref6}, and speech conversion \cite{ref7, ref8}, in real-world conditions. Consequently, there is the need for the development of speech enhancement (SE) to generate enhanced speech with better speech quality and clarity by suppressing background noise components in noisy speech. 

\begin{figure}
  \centering
  \includegraphics[width=0.85\linewidth]{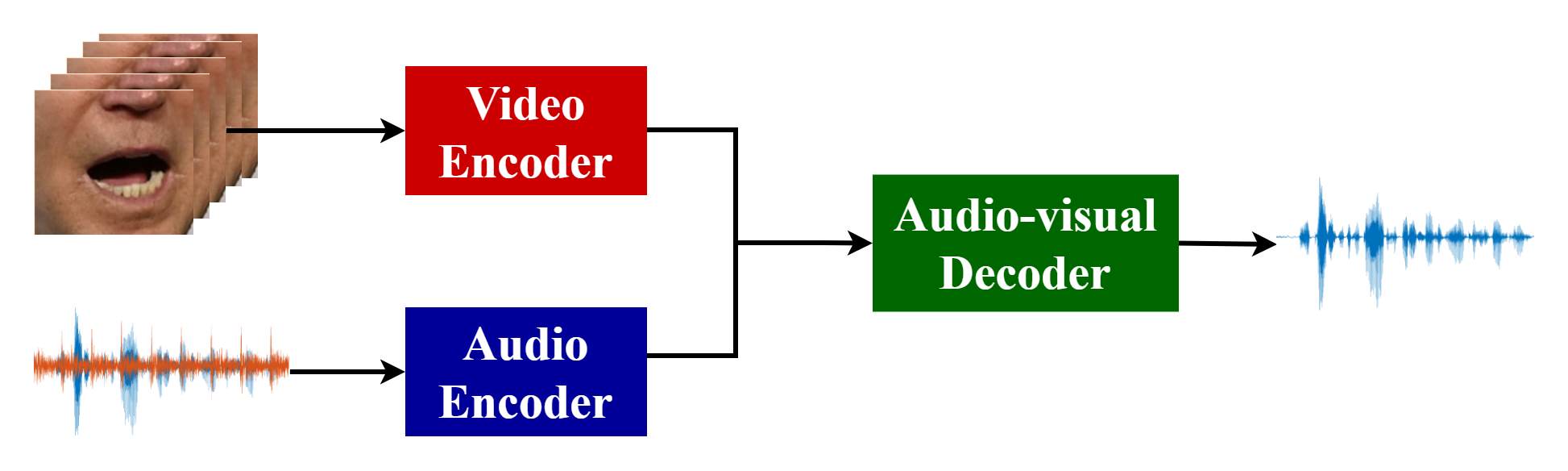}
  \caption{The general deep-learning-based audio-visual speech enhancement system.}
  \label{fig:1}
\end{figure}

Conventional SE approaches, such as spectral subtraction \cite{ref9}, minimum mean squared error (MMSE) estimation \cite{ref10}, Wiener filtering \cite{ref11}, Kalman filter \cite{ref12}, and subspace methods \cite{ref13}, have been extensively studied in the past. Recently, deep learning technologies have been successfully used for SE owing to their superior capability to model complex non-linearity \cite{ref14}. Although these deep learning-based approaches are effective compared to conventional SE approaches, they are prone to a label permutation (or ambiguity) error due to their frame-by-frame or short segment-based processing paradigm \cite{ref15, ref16}. To address this issue, the permutation invariant training using permutation loss criterion is proposed \cite{ref17, ref18}, but the label ambiguity problem still appears in the inference stage, especially for unseen speakers.

Leveraging the visual streams of target speech signals is one of the effective alternatives. Many researchers \cite{ref19, ref20, ref21} have proved that visual cues such as facial/lip movements can supplement acoustic information of the corresponding speaker, helping speech perception, especially in noisy environments. In audio-visual speech enhancement (AVSE) systems, acoustic and visual features are integrated to derive unique characteristics \cite{ref22, ref23, ref24}. In recent AVSE systems, the actual data processing is obtained with deep-learning-based techniques. Generally, acoustic and visual features are processed separately through using two neural network models. After that, the output vectors of these models are fused, often by concatenation, which then is used as input to another deep learning model (as shown in Fig~\ref{fig:1}). This strategy is convenient and easy to be implemented.

\begin{figure*}[t]
  \centering
  \includegraphics[width=0.8\linewidth]{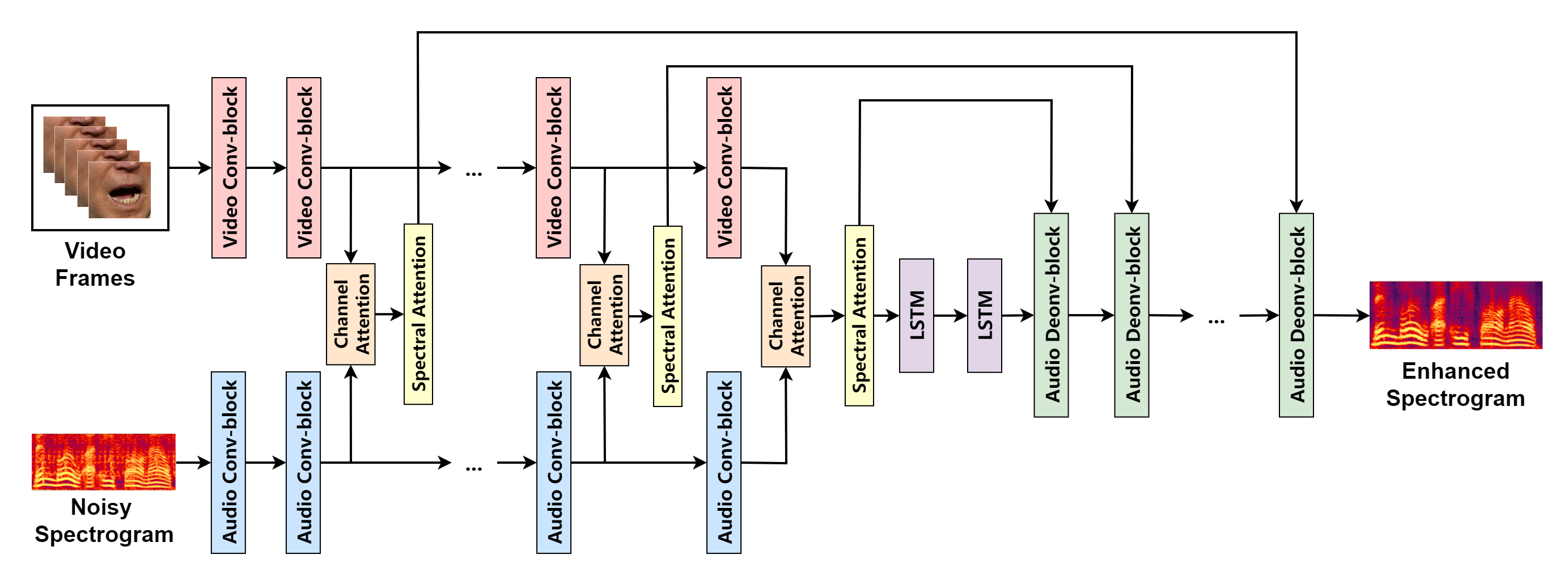}
  \caption{Schematic diagram of the proposed multi-layer feature fusion convolutional network (MFFCN).}
  \label{fig:2}
\end{figure*}
In real-world scenarios, however, this strategy comes with two significant drawbacks, (1) evidence in speech perception suggests that in humans the AV integration occurs at a very early stage \cite{ref25}, the one that processes the two modalities separately and combines them only at a later stage is failed to exploit the the correlation between audio and video at a very early stage, but the disadvantage of early fusion in AVSE systems is that usually the features between the two modalities are inherently different, (2) a simple concatenation does not allow to control how the information from the acoustic and the visual modalities is treated, in consequence, the two modalities may dominate each other, determining a decrease in the system’s performance.

In this letter, we highlight those limitations and tackle the appropriate fusion strategy designing and unbalance between acoustic and visual modalities problems, in AVSE processing. We propose a novel multi-layer feature fusion convolutional network for robust speech enhancement, dubbed as MFFCN, by designing a multi-layer fusion strategy to integrate early fusion and later fusion with a single model, and by designing the channel and spectral attention mechanisms to provide additional flexibility in dealing with different types of information. In particular, the MFFCN separately processes acoustic and visual modalities and fuses audio-visual modalities at each encoder layer to extract AV features, which enjoys the benefit of human auditory perception while avoiding the problem of inherent difference of acoustic and visual modalities during early fusion. The channel and spectral attention mechanisms are inserted in concatenated operation to give the weight to the channels of concatenated features while paying more attention to informative regions of each concatenated feature map. We demonstrate the effectiveness of the proposed network with extensive experiments, achieving large improvements in unconditioned scenarios on GRID \cite{ref26} and TCD-TIMIT \cite{ref27} datasets.

\section{Model Architecture}\label{sec2}

\subsection{Overview}
The diagram of proposed MFFCN is illustrated in Fig~\ref{fig:2}, where its inputs are noisy speech Mel-spectrogram $Y$, and video frames $V$, and it's output is the enhanced speech Mel-spectrogram $S$. The MFFCN follows an encoder-decoder architecture, and consists of encoder part, fusion part, embedding part, and decoder part.

The encoder part of MFFCN is comprised of audio encoder and video encoder.  The audio encoder is designed as a CNN taking spectrogram as input, and each layer of an audio encoder is followed by strided convolutional layer, batch normalization, and Leaky-ReLU for non-linearity. The video encoder is used to process the input face embedding through a number of max-pooling convolutional layers followed by batch normalization, and Leaky-ReLU. Note that the dimension of visual feature vector after convolutional layer has to be the same as the corresponding audio feature vector. In addition, the audio decoder is reversed in the audio encoder part by deconvolutions, followed again by batch normalization and Leaky-ReLU. 

Fusion part designates a merged dimension to implement fusion, and the audio and video streams take the concatenation operation by channel attention for giving the weight to each channel of audio and video feature vectors and are through spectral attention to underscore the informative region of each concatenated feature map. The embedding part is the bottleneck of the encoder-decoder architecture, in which the channel and spectral attention mechanisms are used to process the encoded features and its output is then fed into two LSTM blocks for aggregating temporal contexts. The MFFCN fuse acoustic and visual modalities at early and later stages while separately processing both of modalities, which not only enjoys the benefits of human auditory system but also avoid the the features inherent difference of two modalities.

\subsection{Channel and Spectral Attention Mechanisms}

\begin{figure}[t]
\centering
\subfigure[Channel attention mechanism]{
\label{Fig3.sub.1}
\includegraphics[width=0.25\textwidth]{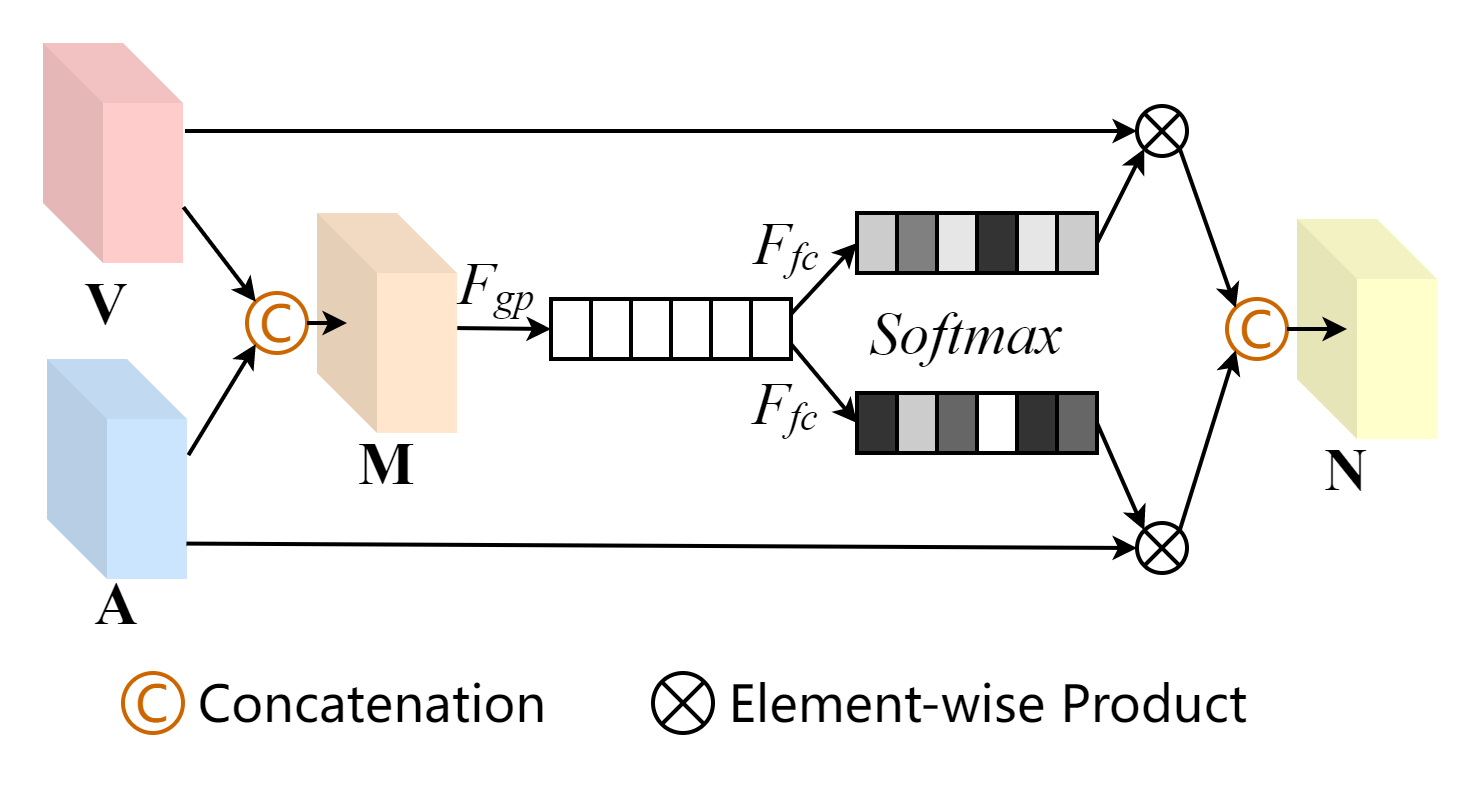}
}
\subfigure[Spectral attention mechanism]{
\label{Fig3.sub.2}
\includegraphics[width=0.2\textwidth]{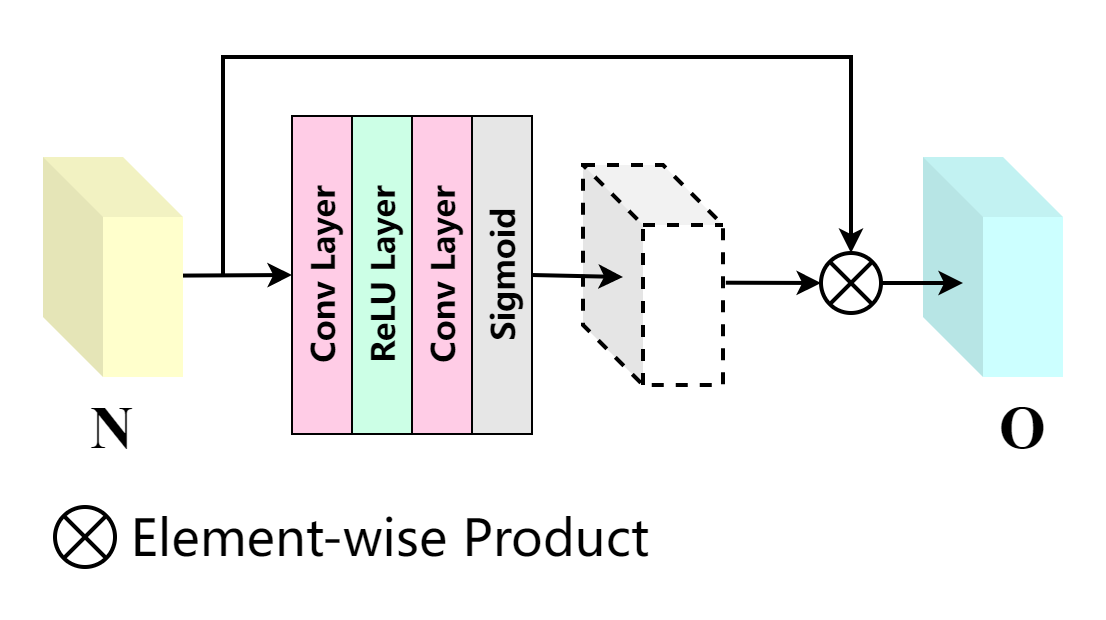}
\centering}
\caption{Schematic diagram of channel and spectral attention mechanisms.}
\label{fig:3}
\end{figure}

\renewcommand{\arraystretch}{0.9}
\begin{table*}[]
\caption{Detailed architecture of the MFFCN encoder. Conv1 denotes the first convolution layer of the MFFCN encoder part.}
\begin{center}
\begin{tabular}{c|cccccccccc}
\hline
                 & Conv1     & Conv2     & Conv3     & Conv4     & Conv5     & Conv6     & Conv7     & Conv8     & Conv9     & Conv10    \\ \hline
Num Filters      &   64      &   64      &   128     &   128     &  256      &  256      &  512      &  512      &  1024     &  1024      \\
Filter Size      & (5, 5)    & (4, 4)    & (4, 4)    & (4, 4)    & (2, 2)    & (2, 2)    & (2, 2)    & (2, 2)    & (2, 2)    & (2, 2)    \\
Stride(audio)    & (2, 2)    & (1, 1)    & (2, 2)    & (1, 1)    & (2, 1)    & (1, 1)    & (2, 1)    & (1, 1)    & (1, 5)    & (1, 1)    \\
MaxPool(video)   & (2, 4)    & (1, 2)    & (2, 2)    & (1, 1)    & (2, 1)    & (1, 1)    & (2, 1)    & (1, 1)    & (1, 5)    & (1, 1)  \\ \hline
\end{tabular}
\end{center}
\label{tab:1}
\end{table*}

Convolutional operation treats the channel-wise and spectral unit-wise feature equally, therefore, extracting AV features by the concatenation operation with convolutional layers is unable to control how the information from the acoustic and the visual modalities is treated. To address this issue, we propose the channel and spectral attention to provide additional flexibility in dealing with different types of information and expand the representational ability of convolution operation.

\subsubsection{Channel Attention Mechanism}
The channel attention mainly concerns that different channel feature have totally different weighted information. Inspired by \cite{ref28, ref29}, we employ the channel-wise selective operation to design channel attention mechanism to perform weighting for different channels with different type of features.

As illustrated in Fig~\ref{Fig3.sub.1}, given acoustic feature $\textbf{A}$ and visual feature $\textbf{V}$ the output of channel attention $\textbf{N}$ is obtained by fusing sub-module in the following four steps:
\begin{align}
    \textbf{M} &= \text{Conv}(\text{Concat}([\textbf{V}, \textbf{A}])), \\
    \textbf{g} &= \text{GlobalPooling}(\textbf{M}),\\
    \textbf{w}_V, \textbf{w}_A &= \text{softmax}([F_{FC}^1(\textbf{g}), F_{FC}^2(\textbf{g})]),\\
    \textbf{N} =& \text{Conv}(\text{Concat}([\textbf{V}\cdot\textbf{w}_V, \textbf{A}\cdot \textbf{w}_A])),
\end{align}
where $F_{FC}^1(\cdot)$ and $F_{FC}^2(\cdot)$ refer to two independent fully-connected layers, ``Conv'' and ``Concate'' represent convolution and concatenation operation, respectively.

To be concrete, we first concatenate the encoded acoustic and visual features from two branches of MFFCN encoder part and apply a convolution block to make the size of concatenation the same as the acoustic or visual features. Then, we adopt a global pooling to produce a global vector $\textbf{g}$ which is used as the guidance of adaptive and accurate channel weighting operation between two modalities. Next, we apply two fully connected layers $F_{FC}^1(\cdot)$ and $F_{FC}^2(\cdot)$ to generate two weight vectors ($\textbf{w}_V$ and $\textbf{w}_A$) to assign the weight for each channel.  In this way, acoustic and visual features can be processed selectively based on their characteristics.

\subsubsection{Spectral Attention Mechanism}
Considering the informative feature distribution is uneven on different feature maps of different modalities, we propose a spectral attention mechanism to make the network pay more attention to informative features, such as lip shape on video modality and speech presence regions on audio modality.

As illustrated in Fig~\ref{Fig3.sub.2}, we directly feed the input $\textbf{N}$ (the output of the channel attention mechanism) into two convolution layers with ReLU and sigmoid activation function to generate the weighting map:
\begin{equation}
    \textbf{w}_N = \sigma(\text{Conv}(\delta(\text{Conv}(\textbf{N})))),
\end{equation}
where $\sigma$ is sigmoid function, and $\delta$ is the ReLU function.

Finally, the input $\textbf{N}$ is element-wise multiplied with the weighting map $\textbf{w}_N$ to obtain the output of spectral attention $\textbf{O}$.

\renewcommand{\arraystretch}{0.95}
\begin{table*}[]
\caption{Performance of trained networks}
\centering
\begin{tabular}{c|cccc|cccc}
\hline
Evaluation metrics             & \multicolumn{4}{c|}{STOI(\%)}                                                          & \multicolumn{4}{c}{PESQ}                                                          \\ \hline
Test SNR                       & \multicolumn{2}{c|}{-5 dB}                           & \multicolumn{2}{c|}{0 dB}       & \multicolumn{2}{c|}{-5 dB}                         & \multicolumn{2}{c}{0 dB}     \\ \hline
Interference                   & Speech         & \multicolumn{1}{c|}{Ambient}        & Speech         & Ambient        & Speech        & \multicolumn{1}{c|}{Ambient}       & Speech        & Ambient       \\ \hline
Unprocessed                    & 57.82          & \multicolumn{1}{c|}{51.44}          & 64.74          & 62.59          & 1.59          & \multicolumn{1}{c|}{1.03}          & 1.66          & 1.24          \\ \hline
Audio-only CRN \cite{ref33}                & 78.26          & \multicolumn{1}{c|}{78.67}          & 80.82          & 83.35          & 2.01          & \multicolumn{1}{c|}{2.19}          & 2.47          & 2.58          \\ \hline
VSE \cite{ref34}                           & 79.94          & \multicolumn{1}{c|}{83.33}          & 84.61          & 87.90          & 2.36          & \multicolumn{1}{c|}{2.41}          & 2.77          & 2.94          \\
AV(SE)$^2$ \cite{ref35}                         & 81.06          & \multicolumn{1}{c|}{85.41}          & 86.17          & 89.44          & 2.42          & \multicolumn{1}{c|}{2.49}          & 2.81          & 2.99          \\
Deep U-Net Early Fusion \cite{ref36}       & 81.23          & \multicolumn{1}{c|}{85.34}          & 87.86          & 90.23          & 2.47          & \multicolumn{1}{c|}{2.53}          & 2.79          & 3.01          \\ \hline
MFFCN (Proposed)               & \textbf{83.24} & \multicolumn{1}{c|}{\textbf{87.82}} & \textbf{90.21} & \textbf{91.46} & \textbf{2.71} & \multicolumn{1}{c|}{\textbf{2.80}} & \textbf{2.97} & \textbf{3.09} \\
MFFCN-w/o channel attention    & 81.36          & \multicolumn{1}{c|}{86.23}          & 88.69          & 89.31          & 2.63          & \multicolumn{1}{c|}{2.75}          & 2.91          & 3.04          \\
MFFCN-w/o spectral attention   & 80.59          & \multicolumn{1}{c|}{85.14}          & 86.36          & 89.23          & 2.56          & \multicolumn{1}{c|}{2.47}          & 2.76          & 3.02          \\
MFFCN-w/o attention mechanisms & 80.47          & \multicolumn{1}{c|}{82.71}          & 85.02          & 88.94          & 2.39          & \multicolumn{1}{c|}{2.47}          & 2.83          & 2.96          \\ \hline
\end{tabular}
\label{tab:2}
\end{table*}

\section{Experimental Setup}\label{sec3}

\subsection{Dataset}
We use two public AV datasets to train our model: the first is GRID \cite{ref26} that consist of video recording where 18 male speakers and 16 female speakers pronounce 1000 sentences each, the second is TCD-TIMIT \cite{ref27}, which comprise 32 male speakers and 30 female speakers around 200 videos each.

We shuffle and split the datasets to training set, validation set, and test set to 24300, 4400, and 1200 utterances, respectively. The noise dataset is self-collected set, containing 25.3 hours ambient noise categorized into 12 types: room, car, instrument, engine, train, human chatting, air-brake, water, street, mic-noise, ring-bell, and music. The speech-noise mixtures in training and validation are generated by randomly selecting utterances from speech dataset and noise dataset and then mixing them up at random SNR between -10dB and 10dB. The evaluation set is generated SNR at 0dB and -5 dB.

\subsection{Training and Network Parameters}
The audio representation is extracted from raw audio waveforms by using Short Time Fourier Transform (STFT) with Hanning window function after resampling the audio signal to 16 kHz. Each frame contains a window of 40 milliseconds, and the frame shift is 160 samples (10 milliseconds). For each speech frame, a log Mel-scale spectrogram is extracted by multiplying the spectrogram via a Mel-scale filter bank, resulting in spectrograms of size 80$\times$20. Visual feature is extracted from the input videos that is re-sampled to 25 frames per second. The video segment processed as input is the sizeof 128$\times$128$\times$5 using the 20 mouth landmarks from 68 facial landmarks suggested by Kazemi et al.\cite{ref30}. For convenience, the processed video segment is zoomed to 80$\times$80$\times$5 by bilinear interpolation algorithm\cite{ref31}. The proposed MFFCN has 10 convolutional layers for each encoder, as shown in Table~\ref{tab:1} and 10 deconvolutional layers for each decoder. The models are trained with the Adam optimizer \cite{ref32} with the learning rate of 0.0002 , and batch size of 8. The mean squared error (MSE) serves as the objective function.

\begin{table}[t]
\centering
\caption{Comparison with different fusion strategies}
\resizebox{0.45\textwidth}{!}{
\begin{tabular}{c|cc|cc}
\hline
\multirow{2}{*}{Fusion Strategy} & \multicolumn{2}{c|}{-5 dB}            & \multicolumn{2}{c}{0 dB}              \\ \cline{2-5} 
                                 & \multicolumn{1}{c}{STOI (\%)} & PESQ & \multicolumn{1}{c}{STOI (\%)} & PESQ \\ \hline
Early Fusion                     & \multicolumn{1}{c}{75.23}     & 2.24 & \multicolumn{1}{c}{81.41}     & 2.66 \\ \hline
Late Fusion                      & \multicolumn{1}{c}{75.01}     & 2.21 & \multicolumn{1}{c}{81.54}     & 2.59 \\ \hline
Intermediate Fusion (VSE)        & \multicolumn{1}{c}{76.68}     & 2.28 & \multicolumn{1}{c}{83.54}     & 2.71 \\ \hline
Intermediate Fusion (AV(SE)$^2$)     & \multicolumn{1}{c}{79.31}     & 2.37 & \multicolumn{1}{c}{84.88}     & 2.79 \\ \hline
Multi-layer Feature Fusion       & \multicolumn{1}{c}{80.47}     & 2.39 & \multicolumn{1}{c}{85.02}     & 2.83 \\ \hline
\end{tabular}}
\label{tab:3}
\end{table}

\section{Results and Analysis}

\subsection{Overall Performance}
The performance of MFFCN is evaluated with the following metrics: Perceptual evaluation of speech quality (PESQ) (from -0.5 to 4.5) \cite{ref37}, and short-time objective intelligibility measure (from 0 to 100($\%$)) \cite{ref38}. Table~\ref{tab:2} presents comprehensive evaluations for four baseline models on untrained noises and untrained speakers, in which ``Speech'' denotes the noise from human speech and ``Ambient'' denotes the ambient noise. In addition, the four elected baseline models are: \textbf{Audio-only CRN}: an audio-only speech enhancement model based on convolutional recurrent network \cite{ref33}.\textbf{VSE}: a middle fusion audio-visual neural network for visual speech enhancement \cite{ref34}. \textbf{AV(SE)$^2$}: an intermediate fusion audio-visual speech enhancement model with several cross-modal squeeze-excitation \cite{ref35}. \textbf{Deep U-Net}: a early fusion audio-visual speech enhancement model with RNN attention blocks \cite{ref36}.

According to Table~\ref{tab:2}, the proposed MFFCN yields the best results in all conditions. Firstly, we compare the audio-only CRN with other 4 models. We observe that even audio-only CRN achieves about 20.44$\%$ to 27.33$\%$ STOI improvement and 0.42 to 1.64 PESQ improvement over unprocessed mixtures, the audio-only CRN is unable to outperform audio-visual SE methods. Secondly, the AV(SE)$^2$ and Deep U-net improves the network performance by adopting different attention mechanisms, and these two models performs better than VSE that is not applying attention operation. This demonstrates the necessity of balancing between acoustic and visual modalities. Finally, our proposed MFFCN consistently outperforms VSE, AV(SE)$^2$, and Deep U-net in all conditions. 

To verify the effect of proposed multi-layer feature fusion strategy, we show the comparison in Table~\ref{tab:3} to investigate the effect of different fusion strategies. Note that (1) \textbf{early fusion} which fuses acoustic and visual modalities at first network layer, \textbf{late fusion} fuse the two modalities at last network layer, (2) \textbf{intermediate fusion (VSE)} that is used in VSE approach and fuse the two modalities at bottleneck, (3) \textbf{intermediate fusion (AV(SE)$^2$)} that is used in AV(SE)$^2$ approach and fuse the two modalities at every decoder layers, (4) \textbf{multi-layer feature fusion} is the proposed fusion strategy. According to Table~\ref{tab:3}, we observe that multi-layer feature fusion significantly performs better than the other 4 fusion strategies, which fully demonstrates the superiority of our proposed fusion strategy.

\subsection{Ablation Study}
\begin{figure}[t]
    \centering
    \subfigure[Noisy]{
    \begin{minipage}[b]{0.135\textwidth}
    \includegraphics[width=1\textwidth]{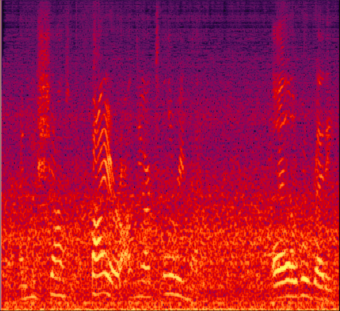}
    \end{minipage}
    }
    \subfigure[w/o C $\&$ S]{
    \begin{minipage}[b]{0.135\textwidth}
    \includegraphics[width=1\textwidth]{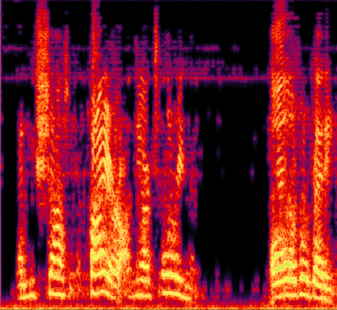}
    \end{minipage}
    }
     \subfigure[w/o S]{
    \begin{minipage}[b]{0.135\textwidth}
    \includegraphics[width=1\textwidth]{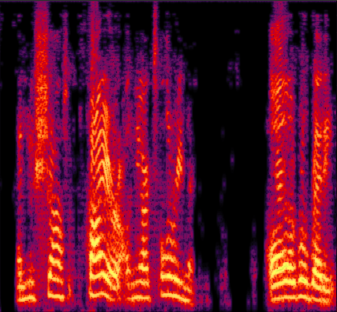}
    \end{minipage}
    }
     \subfigure[w/o C]{
    \begin{minipage}[b]{0.135\textwidth}
    \includegraphics[width=1\textwidth]{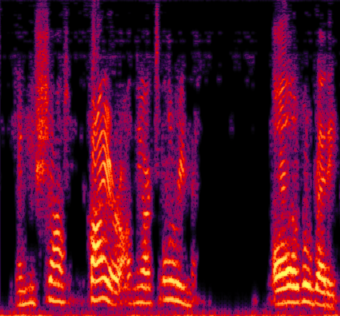}
    \end{minipage}
    }
    \subfigure[Complete]{
    \begin{minipage}[b]{0.135\textwidth}
    \includegraphics[width=1\textwidth]{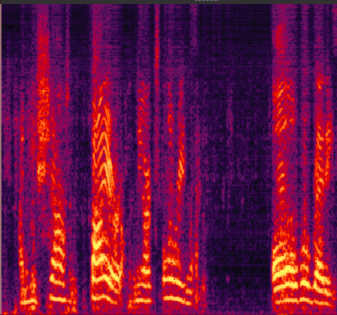}
    \end{minipage}
    }
    \subfigure[Clean]{
    \begin{minipage}[b]{0.135\textwidth}
    \includegraphics[width=1\textwidth]{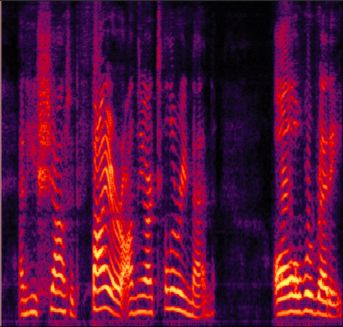}
    \end{minipage}
    }
    \caption{Example of input and enhanced spectra from an example speech utterance. (a) Noisy speech input from test data under the condition of ambient noise at -5 dB. (b) Speech enhanced by MFFCN-w/o channel and spectral attention. (c) Speech enhanced by  MFFCN-w/o channel attention. (d) Speech enhanced by proposed MFFCN-w/o spectral attention. (e) Speech enhanced by MFFCN. (f) Clean speech.} \label{fig:4}
\end{figure}

In this study, we evaluate variants of proposed MFFCN when removing channel attention mechanism, spectral attention mechanism, and two attention mechanisms, respectively. From Table~\ref{tab:2}, we can conclude that the existence of the channel and spectral attention mechanisms does promote the network performance. The channel and spectral attention mechanisms improve 5.11$\%$ and 0.33 on STOI and PESQ, respectively, under the condition of ambient noise at -5 dB. In addition, a single spectral attention mechanism performs better than a single channel attention mechanism. Fig~\ref{fig:4} shows the visualization of noisy speech, MFFCN without channel and spectral attention mechanisms enhanced speech, MFFCN without spectral attention enhanced speech, MFFCN without channel attention enhanced speech, complete MFFCN enhanced speech, and clean speech, respectively. 

\section{Conclusion}
This letter proposes a MFFCN model for AVSE. In particular, the MFFCN adopts multi-layer fusion strategy to fuse AV features layer by layer at encoder phase while separately processing them. In addition, the channel and spectral attention mechanisms are introduced to assign the weight to the channels while paying more attention to informative regions of the fused AV feature maps to avoid the unbalancing between the two modalities. Results show that the proposed model has better performance than recent models on the GRID and TCD-TIMIT datasets.

\end{document}